\documentclass[pre,letterpaper,amsmath,amsthm,
               amssymb,showpacs,twocolumn,floatfix]{revtex4}

\usepackage{graphicx} 
\usepackage{dcolumn} 
\usepackage{array} 
\usepackage{bm} 
\usepackage{latexsym}
\begin{document}

\title{ Segregation in noniteracting binary mixture.}

\author{Filip Krzy\.zewski}
\author{Magdalena A. Za{\l}uska--Kotur}
\affiliation{Institute of Physics, Polish Academy of Sciences,
Al. Lotnik{\'o}w 32/46, 02-668 Warsaw, Poland}

\begin{abstract}
Process of stripe formation is analyzed numerically in a binary mixture. The system consists of particles of two sizes, without any direct mutual interactions. Overlapping of large particles, surrounded by a dense system of smaller particles induces indirect entropy driven interactions between large particles. 
Under an influence of an external driving force the system orders and stripes are formed. 
Mean width of stripes grows logarithmically with time, in contrast to a typical power law temporal increase observed for driven interacting lattice gas systems. 
We describe the mechanism responsible for this behavior and attribute the logarithmic growth to a random walk of large particles in a random potential due to the small ones.   
\end{abstract}

\pacs{68.35.Fx, 61.20.Ja, 64.60Qb, 5.50.+q}
\maketitle

\section{Introduction}
Binary mixtures subject to an external driving force are often found to segregate and form stripes of width increasing with time \cite{Pooley,Ehrhardt,Ciamarra,Mulheran}. Such phenomenon has important practical applications in chemical or pharmaceutical technology. Much theoretical and experimental effort has been devoted to understanding the main mechanisms responsible for this process \cite{Reis,Sanchez,Mullin}. The simplest and best known model of stripe formation under external driving force is a simple lattice gas with attractive nearest neighbor interparticle interactions. It has been known for a long time that such system orders under an influence of the external driving force \cite{Kwan-tai,Katz,Evans}. Recently, this model has been studied intensively \cite{Arapaki,Hurtado,Levine,Gujarti,Furtado,Garrido} in a context of latest experimental results. We describe below the process of stripe formation in a lattice model with two types of particles which do not interact via direct forces. The model is defined in such a way that larger particles, which occupy five lattice sites each, can overlap. Hence they block smaller number of sites, when they are close together, than when they are separated. Presence of smaller particles induces effective interactions between large particles \cite{Frenkel} and the system orders at high enough densities. In the ordering process, due to the presence of smaller particles in the spaces between stripes, large particles realize random walk in a random potential. A jump can occur only when there is enough free space in a chosen direction; the latter being a random event. As a result kinetics of stripe growth in our model is different than typical power law temporal growth known in driven lattice gas \cite{Katz,Hurtado} and a mean stripe width  in the binary mixture studied here increases logarithmically with time.

\begin{figure}

\begin{center}

\includegraphics[width=0.40\textwidth, angle=0]{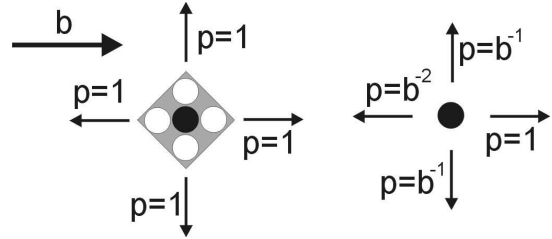}

\caption{\label{Fig. 1. } Jump rules for large (at left) and for small (at right) particles. Shaded area sites can be shared by overlapping large particles. The external bias $b$ affects the jump rates of small particles only. } 

\end{center}

\end{figure}

\begin{figure}

\begin{center}

\includegraphics[width=0.40\textwidth, angle=-90]{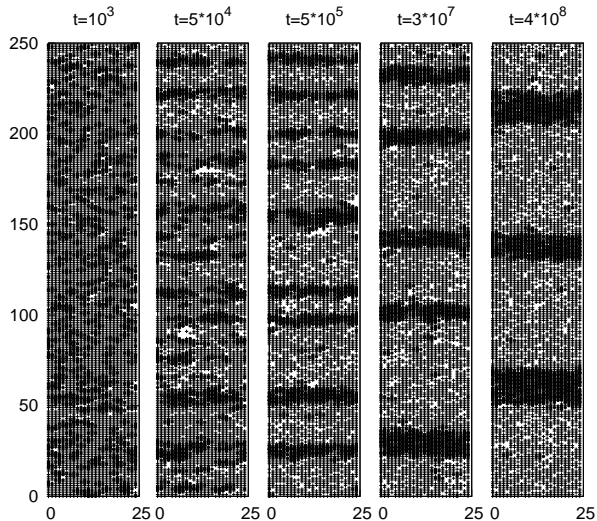}

\caption{\label{Fig. 2. } Successive stages of separation of large (dark) and small (light) particles} 

\end{center}

\end{figure}

\section{Stripe formation in the binary system}   

Stripe formation can be easily observed in a simple lattice gas model with nearest neighbor attractive interactions. It has been shown \cite{ Kwan-tai,Katz,Evans} by means of Monte Carlo simulations that such system orders successively in stripes under an influence of external driving force. When the evolution starts from some random configuration, stripes are formed: initially thin they gradually become thicker.  This process, described and analyzed in detail in Refs \onlinecite{ Arapaki,Hurtado,Levine,Gujarti,Furtado,Garrido} occurs in two stages - stripe formation and then stripe growth. The mean width of stripes grows typically as a power of a simulation time $t^{x}$ (where time is measured in the number of Monte Carlo steps). Typically two different powers $ x=1/3$ and $x= 1/4$ are observed.   
  
Below we present segregation process in a binary mixture of noninteracting particles. The system consists of large objects which can overlap when they occupy neighboring sites and a large number of smaller objects. Large particles occupy area of five sites. The core site in the middle is always occupied by a center of one large particle only, while the other four sites can be shared by other large particles. Small particles occupy one site and do not share it with any other particles. When large particles overlap they leave more free sites for small particles so the number of their possible configurations increases, effectively creating forces between large particles. The system orders forming stripes, when driven by either periodic in time or static external force.  
The large particles move randomly with the same jump probability in each direction. At first, we choose randomly one of four directions. Then occupation of sites in this direction is checked. Each large particle moves only if other particles (large or small) do not block it. Small particles are affected by external, biasing field characterized by a constant $b$. Jump rates of small particles in each direction depend on $b$. When the particle moves in the direction of the bias the jump rate  is $p=1$. When it moves in the direction perpendicular to it then $p=b^{-1}$. For a jump opposite to the field we have  $p=b^{-2}$. In most of our calculations we have used $b=5$. Fig 1. illustrates the jump rules for large and small particles. 

	 There is no direct interaction between particles, but they can partially or fully block sites, which then cannot be occupied by other particles. Large particles can overlap; so they block fewer sites, when they are closer together. When the large particles overlap, then the freed sites become immediately available to small particles. Once they are filled the large ones cannot move away from each other. Therefore, a configuration of closely packed large particles is more probable than other configurations. It is shown in Refs \onlinecite{Frenkel,Gujrati} how to map such a mixture of large and small particles onto an Ising model with nearest-neighbor interactions. Strength of this effective interaction depends on the density of the smaller particles. The interaction is stronger when the density is higher and, eventually, above certain critical density a phase transition occurs. This entropic interaction along with the biasing field causes a formation of stripes parallel to the field. The system structure at different evolution times is shown in Fig. 2. Time for all presented data is measured in a number of Monte Carlo (MC) steps.
 
\begin{figure}

\begin{center}

\includegraphics[width=0.35\textwidth, angle=-90]{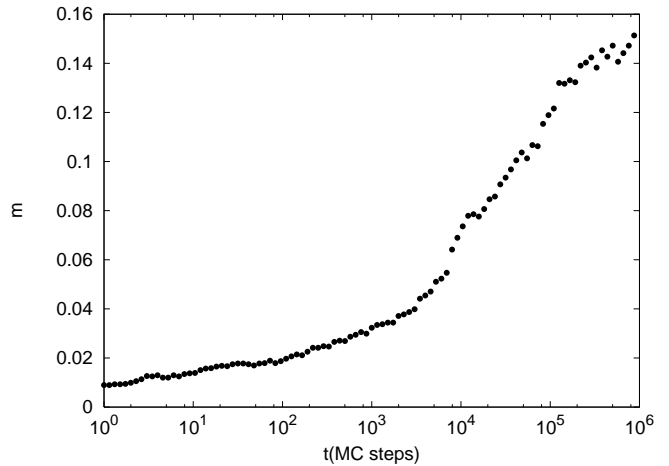}

\caption{\label{Fig. 3. } Number of free sites in the system as a function of time for one sample. System size is ($25 \times 250$) sites populated by   $500$ large and $3972$ small particles. The initial free sites fraction is $0.01$. } 

\end{center}

\end{figure}

\begin{figure}
\includegraphics[width=0.40 \textwidth, angle=0]{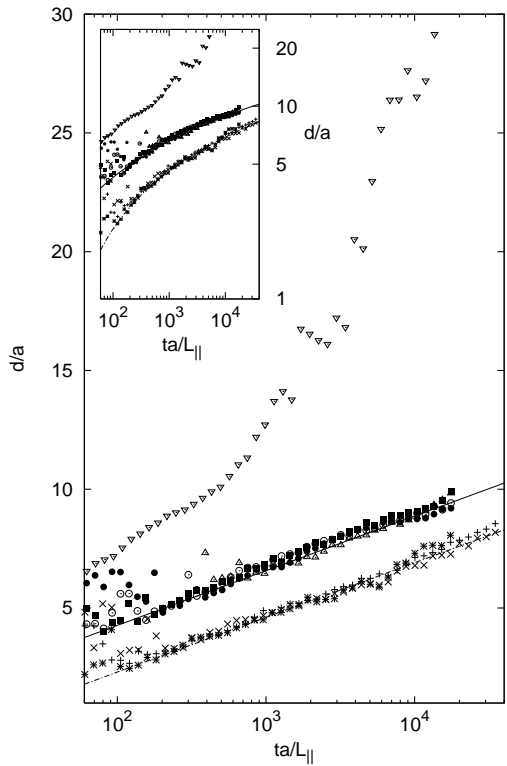}
\caption{\label{Fig. 4. } Average width of stripes $d=\kappa l$ as a function of time, scaled by parameter $\kappa$. See text for explanations.}
\end{figure}	

Increasing fraction of free space is a signature of an ordering process. 
Ratio $m$ of the number of free sites to their total number as a function of time $t$ is plotted in Fig. 3. At $t > 10^4 $ the curve shows a step like structure. Each step lasts a relatively long period of time (note logarithmic time scale in Fig.3) during which the number of stripes does not change. The steps can be observed when there are only few stripes present in the system.
A transition from one step to the other means that the number of stripes is reduced by one.   
Average number of stripes decreases with time $t$ and their average width  $ l(t)$ increases with time, as seen in Fig. 4. We note that, as in  Ref. \cite{Levine} there are two stages of stripe width growth. 
The first stage starts when clusters of large particles coarsen and ends at the time when stripes start to become as long as the system width and a multistripe order becomes clearly visible. The second stage begins with well formed stripes and continues through the process of joining them together. 
These two stages can be seen in Fig. 3. The first stage of stripe formation corresponds to a line of lower slope and the second stage begins when the line bends up and becomes steeper. During the first stage the system looks like in the first left hand side panel in Fig 2, and the second stage is represented in the next four panels. Number of unoccupied sites increases with time.

	 We carried out simulations for various lattice sizes with different width-to-height ratios. 
Results of all simulations were averaged over 100 realizations. Large particles were distributed randomly at the beginning of every run. They could overlap. Then, given large fraction of remaining free sites was occupied randomly by small particles. Then, a biasing field was turned on and the simulation started. We have compared results obtained for closed systems with fixed particle number with those for open systems with fixed chemical potential controlling the particle density. Formation of structures has been observed at a static bias field and at a field with periodically changing orientation. The highest formation rate was noted for a constant field, and it decreased with an increasing frequency of the field variation. The results do not change qualitatively until a frequency of around $1/(5 MC $ steps) is reached, above which stripes stop to form at all. Thus, most of the presented examples here are calculated for a constant in time driving field.

To find an average width of the stripes, a correlation function $f_c(r)$ for a given configuration of particles was calculated. It is defined as

\begin{equation}
f_c(r)=\left\{\begin{array}{ll}
\frac{1}{N}\sum_{[i,j]}n_in_j &\textrm{for $|r|=0$}\\
\frac{1}{2N}\sum_{[i,j]}n_in_j &\textrm{for $|r|\ne0$}
\end{array} \right.
\end{equation}
where $|r|$ is the distance between i-th and j-th site along the direction perpendicular to the external field; $n_i, n_j = 0,1$ are occupations of i-th and j-th site. $n_i=1$ when site $i$ is occupied by one or more large particles and $n_i=0$ when the site is empty.  N is the number of large particles. Sums are over all sites whose coordinates perpendicular to stripes differ by $|r|$. Average width of the stripes is such $|r|$ for which of the correlation function has the first minimum.

Our simulations show that mean width of stripes grows as logarithm of time $l=\log(t/L_{\parallel})$, where $L_{\parallel}$ is the system width, parallel to stripes ($L_{\parallel}=25$ in Fig. 2). The character of growth does not depend on the system size. This is illustrated in the Fig. 4, where seven data sets, are plotted in two groups for the binary mixture and top plot is for a one component, interacting driven lattice gas. The main panel shows the mean width of stripes $l$ as a function of $\log(t)$. 
Three lower data sets, plotted by $ \divideontimes, + $ and $\times $ , represent results for systems of constant number of smaller particles, and four data sets in the middle, plotted by $\blacksquare, \circ, \bullet$  and $\vartriangle$ represent results for open systems, with fixed chemical potential.
For all these plots stripe width $l$ was rescaled by a parameter $\kappa$ chosen in such a way that each data set lies on one line $\log[t/(\alpha L_{\parallel})] $ with $\alpha=10$ for closed and $\alpha=1.4$ for open systems. Scaling parameters for closed systems are $ \kappa= 0.4,0.5 $ and $0.7$ for the ratio $r=L_{\parallel}/L_{\perp}$ equal, respectively to $2, 4$ and $10$. System sizes are ($ 50 \times 100$), ($25 \times 100$), and ($ 25 \times 100$) and number of large particles $N=600, 300$ and $600$. For open systems $\kappa=0.833,1,1.66$ and $1.25$, respectively. Size of each system is ($25 \times 100$), $N=600$ and  $80\%$,  $85\%$,  $90\%$ and  $95\%$ sites are occupied, respectively.  The upper set printed by triangles $\triangledown$  represents a stripe growth for one component interacting driven  system at a temperature $0.8 {T_c}^{\infty}$, where ${T_c}^{\infty}=3.18 J/k_{B}$ (where $J$ is interaction strength) and jump probabilities:  $p=1$ in the direction of field, $p=0$ in the opposite direction, and  $p=\exp(- c J/T )$ in the direction perpendicular to the field, where $c$ is a number of nearest neighbors. 

We can now compare results for a one-component interacting driven system with those for a non-interacting binary mixture. Stripe formation is observed in both cases. It can be seen, however, that time dependence of these processes has a different character.
In contrast to typically noted power law in time stripe width growth for one component interacting driven systems, in binary mixtures with entropic interactions we typically observe slower logarithmic time dependence.
Ordering in the binary system happens due to the indirect, effective attractive interaction between particles. Strength of this entropic interaction decreases with increasing number of free sites \cite{Frenkel}. In the closed system with a fixed particle number the available free space expands with time (Fig. 3) and the effective interactions weaken. In order to check if this weakening has influence onto the character of stripe growth, we studied also open systems with varying small particle number and where mean number of free sites was controlled by an external potential.  As illustrated in Fig. 4, the time dependence in open system has the same logarithmic character observed for the closed system. We see that the decreasing with time interaction strength is not the main reason for the type of observed temporal width growth. 

Specific aspect of the system studied here is the existence of particles of two types. In order to cross an interstripe distance, large particle have to find their way between densely packed set of small ones. To execute a jump, the large particle has to wait until a passing stream of small particles creates a hole, large enough to fit in. As a waiting time for a jump in such case varies from one event to the next,
we can treat such process as a random walk in a random potential.  In the next section we show that the logarithmic character of the temporal stripe width growth can be explained by such a description of large particles kinetics.

\section{Mechanism of stripe growth}

Stripe growth is an anisotropic process that takes place in the driven systems. The main course of growth happens along the direction perpendicular to stripes. Existence of the second dimension controls relative probabilities of several mechanisms that compete in the stripe growth process. This process in one component system has been analyzed and explained in details in Refs \onlinecite{Levine} and \onlinecite{Hurtado}. Ref. \onlinecite{Hurtado} describes two different competing mechanisms: evaporation/condensation of particles from the surface of the stripe and diffusion of particles/holes between interfaces. The former one leads to $l \sim t^{1/4}$ and can be observed at earlier times or for shorter systems, whereas the latter leads to the $ l \sim t^{1/3}$ growth and is activated at later stages of stripe formation or in longer systems.

Let us consider a one component system with particles attracting each other. System orders under influence of a static bias field, initially forming many thin stripes. When the process continues some of stripes disintegrate while the remaining ones become thicker. Stripe extinction is a process consisting random actions of a single particle: the particle evaporates first from the stripe wall, then it walks randomly in an empty space until it readsorbs at the same or the other wall. The process continues until one of stripes disintegrates.
Decay of one of two neighboring stripes is a problem similar to that of the gambler ruin. We are not asking, however, a standard question about the probability of a ruin. Instead, we are interested in the mean time of ultimate decay of the first or the one of two neighboring stripes. This time is proportional to the mean time of evaporation of one particle row across one stripe. Number of particles in such row is equal to the width of the stripe and fluctuates as particles escape from and stick to the domain walls. Emergence of a fluctuation of size $l$ means that row of such length disappears. Mean time of such an event scales as $l^2$. Fluctuations occur independently in each row, so the time in which the entire stripe disappears is proportional to the number of rows in one stripe, $L_{\parallel}$ and to the time $\tau_0$ needed for a single particle to pass the distance from one stripe to another. Thus we have
\begin{equation}
\label{tau}
\tau=\nu L_{\parallel}\tau_0 l^2 ,
\end{equation}
where $\nu$ is time scale parameter. 

Time $\tau_0$ is mean first-passage time of a distance between stripes by a particle. In a general case of an inhomogeneous potential the first-passage time is given by \cite{Kehr,Noskowicz}
\begin{equation}
\label{first_passage}
\tau_0=\sum_{n=0}^{s-1}\frac{1}{p_n}\sum_{k=n}^{s-1}\prod_{j=n}^k \frac{q_j}{p_j} 
\end{equation} 
where $s=l(1-\rho)/\rho$ is the distance between stripes, $\rho$ a density of large particles in the system, $p_j$ is jump rate from site $j$ in the direction pointing from the initial site $0$ to the final site $s$, and $q_j$ is jump rate in the opposite direction. For the one component system we can assume that $p_j=q_j$ in the Eq. (\ref{first_passage}) and that $p_j$ are the same for all $j=1,...s$ except when $j=0$ for a jump originating at a site neighboring to the stripe. The rate $p_0$ is a probability rate for a particle jumping out of other particles. The particles attract each other, so this rate is smaller than all others:  $p_0<p_1$. The interaction and so the value of $p_0$ depend on the number of neighboring particles. We can write 
\begin{equation}
\label{tau0}
\tau_0 \sim \frac{s}{p_0}+\frac{s^2}{p_1}
\end{equation}
and treat $p_0$ as an effective rate averaged over many jumps. The above formula is correct when the density of particles between stripes is low, not higher than one free particle per row. If $s>{p_1}/{p_0}$ then the second term of (\ref{tau0}) dominates. In this case, however, the particle density is higher, than one particle per row so proper expression for the time $\tau_0$ is obtained by dividing by a number of particles that reach the wall per time unit. This number is proportional to the distance $s$, so for $s>{p_1}/{p_0}$  we get 
\begin{equation}
\label{tau0b}
\tau_0 \sim \frac{s}{p_1}.
\end{equation}
Equation (\ref{tau0b}) is valid if the density of particles between stripes is higher than one particle per row but is still quite low. For higher densities, however, pair interactions in the empty space start to play a role, causing the entire process to slow down.
Using Eq. (\ref{tau}) we obtain    
\begin{equation}
\frac{d {l}}{d t}=\frac{l}{\tau}=\frac{1}{L_{\parallel} l \tau_{0}}.
\end{equation}
Solution of this equation for $\tau_0$ given in (\ref{tau0}) is
\begin{eqnarray}
\frac{(1-\rho){l}^3}{3p_0\rho}+\frac{{l}^4 (1-\rho)^2}{4 p_1 \rho^2} \sim \frac{t}{L_{\parallel}},
\end{eqnarray}
the exponent of the power law growth of $l$ changes between $1/3$ and $1/4$. When (\ref{tau0b}) is used, we get
\begin{eqnarray}
l \sim ({\frac{t}{L_{\parallel}}})^{1/3}. 
\end{eqnarray}

i.e. the power law time dependence with a single exponent $x =1/3$ \cite{Hurtado}.
For higher temperatures when particle density between stripes becomes higher, various values of exponent $x$, usually smaller than $1/3$, are observed. Still, $l \sim t^{1/3}$ is a dominant behavior for wide range of temperatures and system geometry parameters.
 
 When the system consists of two different types of particles, random walk from one stripe to another is not free. Each particle has to wait until there is enough space for it jump. We can treat the process of particle motion in a dense medium as a random walk in a random potential landscape. A jump to the left with rate $p_l$ and jump to the right with rate $p_r$ are in this approach independent events, occurring according to the same probability distribution. Such a model leads to the following expression for the mean first-passage time \cite{Kehr}
\begin{equation}
\label{taut}
\tau_0 \sim \frac{2\gamma (\gamma^s-1)}{(\gamma-1)^2} \sim e^{\lambda l}
\end{equation}
where 
\begin{equation}
\gamma=\langle p_l \rangle \langle \frac{1}{p_r} \rangle   > 1
\end{equation}
with $\langle \rangle $ being an average over random variable realizations. Thus all linear in $s$ terms in expression (\ref{taut}) are for large $s$ irrelevant and we get $\lambda=\log(\gamma) (1-\rho)/\rho $. Using now (\ref{tau}) and (\ref{taut}) we get the following equation 
\begin{equation}
\frac{d l}{dt} \sim \frac{e^{-\lambda l}}{L_{\parallel} l}.
\end{equation}
Its solution  for large $l$ and $t$ can be written as
\begin{equation}
l \sim \log(t/L_{||})
\end{equation}
and, indeed such character of the time dependence is observed in Fig 4 for binary systems. It can be seen in the inset of Fig 4., that power law cannot be fitted to the data sets for binary mixtures.
The character of stripe growth is the same for closed system, where number of free sites increases as it is for an open system with constant density of small particles controlled by external potential.

\section{Summary}
We have investigated binary mixture system driven by an external force. Particles in this binary system do not interact with each other directly but they effectively do so via indirect entropy interaction. The system orders forming stripes, similarly like in driven single component system with attractive forces.

The existence of two different particle types leads to the logarithmic temporal growth of the mean stripe width. Such time dependence is slower than the power law temporal growth in an interacting one component system. In binary systems large particles travel among densely packed small particles, which effectively slow down their wandering. We have attributed the logarithmic growth process to a random walk of large particles in an effectively random potential.

\begin{acknowledgements}
This work was supported by Poland Ministry of Sciences and Higher Education Grant No N202 042 32/1171
\end{acknowledgements}

\end{document}